\documentstyle[12pt]{article}

\begin{document}

\author{R. Vilela Mendes \\
Grupo de F\'\i sica-Matem\'atica\\
Complexo Interdisciplinar, Univ. de Lisboa \\
Av. Gama Pinto, 2, 1699 Lisboa Codex Portugal\\
e-mail: vilela@alf4.cii.fc.ul.pt}
\title{Gluon condensate and the vacuum structure of QCD }
\date{}
\maketitle

\begin{abstract}
Phenomenological evidence and analytic approximations to the QCD\ ground
state suggest a complex gluon condensate structure. Exclusion of elementary
fermion excitations by the generation of infinite mass corrections is a
consequence. In addition the existence of vacuum condensates in unbroken
non-abelian gauge theories, endows SU(3) and higher order groups with a
non-trivial structure in the manifold of possible vacuum solutions, which is
not present in SU(2). This may be related to the existence of particle
generations.
\end{abstract}

\section{Vacuum condensates in QCD. Phenomenological and theoretical evidence
}

There is now ample phenomenological evidence for the existence of a
non-trivial structure in the QCD vacuum, containing both quark\cite{GellMann}
and gluon\cite{Shifman1} condensates. A good description of many hadronic
quantities is obtained in the framework of the QCD sum rules\cite{Shifman2},
using as input the vacuum expectation values $\left\langle \overline{q}%
q\right\rangle $ and $\left\langle F_a^{\mu \nu }F_{\mu \nu }^a\right\rangle 
$.

Analytical approximations to the QCD\ ground state also provide theoretical
evidence for the existence of the condensates and, in addition, supply some
additional information on the nature of the condensates. In Ref.\cite
{Vilela1}, for example, an exact path-integral representation of the ground
state is used to obtain a systematic expansion in which even the leading
term contains non-perturbative information. For the gluon sector of QCD (in
the temporal gauge),\ the leading term is 
\begin{equation}
\label{1.3}\Psi _0\left\{ A\right\} =\exp \left( -\frac 12\int
d^3xB_k^\alpha (x)\left( \frac 1{\sqrt{R\left( A(x)\right) R\left(
A(x)\right) }}\right) _{kk^{^{\prime }}}^{\alpha \alpha ^{^{\prime
}}}B_{k^{^{\prime }}}^{\alpha ^{^{\prime }}}(x)\right) 
\end{equation}
with 
\begin{equation}
\label{1.4}B_i^\alpha =\epsilon _{ijk}\left( \partial _jA_k^\alpha -\frac
g2f_{\alpha \beta \gamma }A_j^\beta A_k^\gamma \right) 
\end{equation}
\begin{equation}
\label{1.5}R(A)_{nn^{^{\prime }}}^{\alpha \alpha ^{^{\prime }}}=\epsilon
_{nmn^{^{\prime }}}\left( \partial _m\delta _{\alpha \alpha ^{^{\prime
}}}-gf_{\alpha \gamma \alpha ^{^{\prime }}}A_m^\gamma \right) 
\end{equation}
In the long-wavelength limit, the $\Psi _0\left\{ A\right\} $ of Eq.(\ref
{1.3}) bears some resemblance to the ansatz proposed by Greensite\cite
{Greensite1}, however, the power dependence on the chromomagnetic fields is
different. By considering either fast varying potentials $A_k^\alpha $ or
constant field configurations, on sees that (\ref{1.3}) interpolates nicely
between an abelian-type vacuum for the high frequencies and a configuration
of random magnetic fluxes for the constant fields.

The non-perturbative nature of this analytical approximation to the (gluon
sector) vacuum is made apparent when, for example, one considers the effect
of long-wavelength (constant $A_k^\alpha $) field fluctuations on the
effective mass and propagation of elementary fermions. This was already
briefly mentioned in \cite{Vilela1} and is described in some detail in
Sect.2.

In addition some information may be gathered from (\ref{1.3}) concerning the
group-theoretical structure of the QCD vacuum. This, however, is likely to
be more general than the approximation described in Eq.(\ref{1.3}) and to
depend only on the octet structure of the operators that appear in the
composite condensates $\left\langle A_a^\mu A_{a\mu }\right\rangle $ and $%
\left\langle F_a^{\mu \nu }F_{a\mu \nu }\right\rangle $. A gluonic ground
state, of the type of Eq.(\ref{1.3}), defines a probability distribution for
chromomagnetic field fluctuations around zero mean in such a way that 
\begin{equation}
\label{1.6}\left\langle A_a^\mu \right\rangle =\left\langle F_a^{\mu \nu
}\right\rangle =0 
\end{equation}
as required by Lorentz and SU(3) invariance, while quantities like $%
\left\langle A_a^\mu A_{a\mu }\right\rangle $ and $\left\langle F_a^{\mu \nu
}F_{a\mu \nu }\right\rangle $ may be different from zero. The integration
measure that controls the ground state fluctuations is $\left| \Psi
_0\{A\}\right| ^2\prod_{x,k,\alpha }dA_k^\alpha (x)$. The domain of
integration of this measure is the space of all possible ground state
fluctuations, which should be defined in such a way as to preserve the
symmetries of the theory. In the temporal gauge, one may, by space
rotations, transform the gauge potential $A_a^\mu (x)$ at each point $x$
into a triplet of SU(n) orthogonal directions. Then, the directions to
include in the domain of integration are those that can be obtained from
this triplet by arbitrary SU(n) rotations. If the gauge group is SU(2),
these three orthogonal directions in SU(2) space may, by SU(2)
transformations, be brought to any other orthogonal directions. This is
because the Lie algebra of SU(2) coincides with the Lie algebra of SO(3). It
means that, to implement full SU(2)-symmetry, all possible directions have
to be included in the domain of integration. Hence the vacuum is unique.

For SU(3), however, the situation is different. By SU(3) transformations one
cannot rotate an octet vector and, even less, a triplet of octets, to all
possible directions in $R^8$. This is the well-known fact that the SU(3)
group has a non-trivial structure in the octet space\cite{Michel}. To
preserve SU(3) symmetry it suffices to include in the vacuum the fluctuating
directions that can be reached from a particular one by SU(3)
transformations. Denoting by $VO_8$ the space of orthogonal frames in $R^8$
we conclude that there are as many non-equivalent ways to construct ground
states compatible with SU(3) invariance as there are SU(3) orbits in $VO_8$.
Conversely, if one lacks a specific reason to prefer any particular orbit,
one might instead say that the gluonic vacuum has a larger SO(8) symmetry,
because SO(8) is the smallest group that rotates any three-frame in $VO_8$
into any other. Therefore the possible vacuum structures correspond to
representations of the homogeneous coset space SO(8)/SU(3). This reasoning
holds independently of the particular form of the ground state measure
density $\left| \Psi _0\{A\}\right| ^2$ . The consequences of the SO(8)
structure of a QCD\ vacuum with gluon condensate are explored in Sect.3.

Greensite\cite{Greensite1} and Feynman\cite{Feynman}, inspired by the form
of the abelian ground state and the requirements of gauge invariance, have
conjectured a form 
\begin{equation}
\label{1.7}\Psi _0\left\{ A\right\} =\exp \left( -\int d^3xd^3y\textnormal{Tr}%
\left( B(x)\bullet S_{xy}\bullet B(y)\bullet S_{yx}\right) f(\left|
x-y\right| )\right) 
\end{equation}
for the (gluonic) QCD ground state. $S_{xy}$ is the gauging factor%
$$
S_{xy}=\exp \left( -\frac i2\int_x^yA_i(\xi )d\xi ^i\right) 
$$
and for the kernel $f(\left| x-y\right| )$, which contains all the
non-trivial dynamical information, these authors have discussed a few
requirements. However comparing Eq.(\ref{1.7}) with the leading
path-integral approximation of Eq.(\ref{1.3}) the conclusion is that a
simple coordinate dependence for the kernel is unlikely. Instead one obtains
a complex dependence on the dynamical variables which seems difficult to
guess from qualitative considerations. One also notices that Eq.(\ref{1.3})
may be written as 
\begin{equation}
\label{1.8}\Psi _0\left\{ A\right\} =\exp \left( -\frac 12\int d^3x\xi
(x)\bullet \xi (x)\right) 
\end{equation}
where 
\begin{equation}
\label{1.9}
\begin{array}{ccl}
\xi _k^\alpha (x) & = & \left( \left( \frac 1{R\left( A(x)\right) R\left(
A(x)\right) }\right) ^{1/4}\right) _{kk^{^{\prime }}}^{\alpha \alpha
^{^{\prime }}}B_{k^{^{\prime }}}^{\alpha ^{^{\prime }}}(x) \\  
& = & \frac 1{\pi \sqrt{2}}\int_0^\infty d\lambda \lambda ^{-1/4}\frac
1{\lambda +R\bullet R}\bullet B(x) 
\end{array}
\end{equation}
the last equality being obtained from a standard representation\cite{Yosida}
for fractional powers of positive operators. Eq.(\ref{1.9}) puts into
evidence the highly non-local nature of the effective QCD coordinates.

\section{Gluon condensate and the effective mass of elementary fermion
excitations}

As will be seen later on, the nature of the gauge group determines the
multiplicity of vacuum configurations, the SU(2) and SU(3) groups behaving
differently in this aspect. However, dynamical features like the coupling
constant dependence of the condensates and their effect on the propagation
of elementary excitations are expected to depend mostly on the non-Abelian
character and not so much on the order of the gauge group. Therefore, for
simplicity, calculations in this section will be carried out for the SU(2)
group.

In the Schr\"odinger formulation, the Hamiltonian involves products of
fields and functional derivatives at the same point. Regularization of these
ill-defined quantities is needed. The simplest way is to use a lattice
cut-off, with the ultraviolet (continuum) limit obtained when the lattice
spacing $a\rightarrow 0$. As shown in \cite{Vilela1} for the quantum theory
constructed from (\ref{1.3}) the existence of a finite mass gap implies a
running coupling constant behavior $g^2(a)\sim \left| \frac c{\log a}\right| 
$. Therefore $g(a)\rightarrow 0$ when $a\rightarrow 0$ which justifies the
use of the small noise Wentzell-Freidlin technique to analyze the mass-gap.
In the lattice regularization one makes the substitutions 
\begin{equation}
\label{2.1}gaA_i^\alpha (x)\rightarrow \theta ^\alpha (x,x+\widehat{i}%
)=\theta _i^\alpha (x) 
\end{equation}
\begin{equation}
\label{2.2}ga^2B_i^\alpha (x)\rightarrow \beta _i^\alpha (x)=\frac 14\gamma
_{ijk}\left( \theta ^\alpha (x+\widehat{j},x+\widehat{j}+\widehat{k})-\frac
12f_{\alpha \beta \gamma }\theta ^\beta (x,x+\widehat{j})\theta ^\gamma (x,x+%
\widehat{k})\right) 
\end{equation}
\begin{equation}
\label{2.3}\left( D_i\right) _{\alpha \beta }v^\beta (x)\rightarrow \frac
1a\left( {\cal D}_i\right) _{\alpha \beta }v^\beta (x)=\frac 1a\left\{ \frac
12\left( v^\alpha (x+\widehat{i})-v^\alpha (x-\widehat{i})\right) -f_{\alpha
\gamma \beta }\theta ^\gamma (x,x+\widehat{i})v^\beta (x)\right\} 
\end{equation}
where $\gamma _{ijk}=($sign$i)($sign$j)($sign$k)\epsilon _{|i||j||k|}$ and $%
\widehat{i}$ denotes the unit lattice vector along the $i$-direction. Then (%
\ref{1.3}) becomes 
\begin{equation}
\label{2.4}\Psi _0\left\{ \theta \right\} =\exp \left( -\frac 1{2\pi
g^2}\sum_x\int_0^\infty d\lambda \lambda ^{-\frac 12}\beta _k^\alpha
(x)\left( \frac 1{\lambda +{\cal R}_x\bullet {\cal R}_x}\right)
_{kk^{^{\prime }}}^{\alpha \gamma }\beta _{k^{^{\prime }}}^\gamma (x)\right) 
\end{equation}
with ${\cal R}\left( \theta \right) _{nn^{^{\prime }}}^{\alpha \alpha
^{^{\prime }}}=\epsilon _{nmn^{^{\prime }}}{\cal D}_m\left( \theta \right)
^{\alpha \alpha ^{^{\prime }}}$ and a standard integral representation\cite
{Yosida} was used for the fractional power of the operator.

To study the long-wavelength contribution to the vacuum condensates,
consider the case of constant gauge potentials. Because $M_{ij}=\sum_\alpha
\theta _i^\alpha \theta _j^\alpha $ is a symmetric matrix it may
diagonalized by a space rotation and in the new coordinates the three SU(2)
vectors $\theta _1^\alpha $ $\theta _2^\alpha $ and $\theta _3^\alpha $ are
orthogonal. Without loosing generality, SU(2) coordinates may be chosen such
that 
\begin{equation}
\label{2.5}\theta _1^\alpha =\left( a_1,0,0\right) \textnormal{ ; }\theta
_2^\alpha =\left( 0,a_2,0\right) \textnormal{ ; }\theta _3^\alpha =\left(
0,0,a_3\right) 
\end{equation}
\begin{equation}
\label{2.6}\beta _1^\alpha =\left( -a_2a_3,0,0\right) \textnormal{ ; }\beta
_2^\alpha =\left( 0,-a_3a_1,0\right) \textnormal{ ; }\beta _3^\alpha =\left(
0,0,-a_1a_2\right) 
\end{equation}
Then (\ref{2.4}) becomes $\Psi _0\left\{ \theta \right\} =\exp \left( \sigma
\left( a_1,a_2,a_3\right) \right) $ with 
\begin{equation}
\label{2.7}
\begin{array}{ccr}
\sigma \left( a_1,a_2,a_3\right)  & = & -\frac N{2\pi g^2}\int_0^\infty
d\lambda \lambda ^{-\frac 12}\{\left( a_1,a_2,a_3\right) ^2\left(
a_1^2+a_2^2+a_3^2\right)  \\  
&  & +\lambda (a_1^4(a_2^2+a_3^2)+a_2^4(a_1^2+a_3^2)+a_3^4(a_1^2+a_2^2)) \\  
&  & +\lambda ^2(a_2^2a_3^2+a_1^2a_3^2+a_1^2a_2^2)\} \\  
&  & \times \{4(a_1a_2a_3)^2+\lambda (\lambda +a_1^2+a_2^2+a_3^2)^2\}^{-1}
\end{array}
\end{equation}
N is the number of sites in the regularizing lattice. The state $\Psi
_0\left\{ \theta \right\} =\exp \left( \sigma \left( a_1,a_2,a_3\right)
\right) $, as it stands, is not normalizable. This has two origins. First,
as one sees from (\ref{2.6}), there still is a gauge freedom on the planes
where one of the arguments vanish. This is corrected by integrating not on $%
da_1da_2da_3$ but on the components of the chromomagnetic field. This is
equivalent to introduce the Jacobian factor $\sqrt{a_1^2a_2^2a_3^2}$. Even
then the state is not normalizable unless on restricts the large potential
fluctuations. This is done by multiplying the integration measure by $\exp
\left\{ -\mu \left( a_1^2+a_2^2+a_3^2\right) \right\} $ with $\lim _{\mu
\rightarrow 0}$ being taken in the end. Expectation values of operators are
therefore obtained from 
\begin{equation}
\label{2.7a}<O>=\lim _{\mu \rightarrow 0}\frac{\int \sqrt{a_1^2a_2^2a_3^2}%
e^{-\mu \left( a_1^2+a_2^2+a_3^2\right) }da_1da_2da_3\Psi _0\left\{ \theta
\right\} \bullet O\bullet \Psi _0\left\{ \theta \right\} }{\int \sqrt{%
a_1^2a_2^2a_3^2}e^{-\mu \left( a_1^2+a_2^2+a_3^2\right) }da_1da_2da_3\Psi
_0^2\left\{ \theta \right\} }
\end{equation}
and because $g(a)\rightarrow 0$ when $a\rightarrow 0$, these integrals may
be evaluated by asymptotic expansion methods, namely the Laplace method. The
minimum of $\sigma (a_1,a_2,a_3)$ is zero and is obtained when any two of
its arguments vanish. Fixing each turn one of the arguments and applying the
Laplace method in the plane of the other two arguments, one obtains three
similar contributions. For example for fixed $a_3$ one computes the second
derivatives in the plane $(a_1,a_2)$%
\begin{equation}
\label{2.8}
\begin{array}{rclcl}
\sigma (0,0,a_3) & = & 0 &  &  \\ 
\frac \partial {\partial a_1}\sigma (0,0,a_3) & = & \frac \partial {\partial
a_2}\sigma (0,0,a_3) & = & 0 \\ 
\frac{\partial ^2}{\partial a_1^2}\sigma (0,0,a_3) & = & \frac{\partial ^2}{%
\partial a_2^2}\sigma (0,0,a_3) & = & 2\pi 
\sqrt{a_3^2} \\ \frac{\partial ^2}{\partial a_1\partial a_2}\sigma (0,0,a_3)
& = & 0 &  & 
\end{array}
\end{equation}
Using $\int_0^\infty z^k\exp (-\alpha z^2)dz=\Gamma (\frac{k+1}2)/(2\alpha ^{%
\frac{k+1}2})$ and the generalized Laplace method one obtains for the
normalization integral at small $g$%
\begin{equation}
\label{2.9}
\begin{array}{lcc}
\int d\nu (a_1,a_2,a_3)\Psi _0^2\left\{ \theta \right\} = &  &  \\ 
3\int_{-\infty }^\infty da_3\left( \frac 1{\frac N{2g^2}\left| a_3\right|
+\mu }\right) ^2\left| a_3\right| e^{-\mu a_3^2} & \sim  & -9\left( \frac{%
2g^2}N\right) ^2\ln \mu 
\end{array}
\end{equation}
where I have denoted by $d\nu (a_1,a_2,a_3)$ the measure 
\begin{equation}
\label{2.10}d\nu (a_1,a_2,a_3)=\sqrt{a_1^2a_2^2a_3^2}e^{-\mu \left(
a_1^2+a_2^2+a_3^2\right) }da_1da_2da_3
\end{equation}
and the last expression in (\ref{2.9}) is the asymptotic behavior for small $%
\mu $. Likewise 
\begin{equation}
\label{2.11}
\begin{array}{rcl}
\int d\nu (a_1,a_2,a_3)\Psi _0^2\left\{ \theta \right\} \left(
a_1^2+a_2^2+a_3^2\right)  & \sim  & 3\left( 
\frac{2g^2}N\right) ^2\frac 1\mu  \\ \int d\nu (a_1,a_2,a_3)\Psi _0^2\left\{
\theta \right\} \left( a_1+a_2+a_3\right) ^2 & \sim  & 3\left( 
\frac{2g^2}N\right) ^2\frac 1\mu  \\ \int d\nu (a_1,a_2,a_3)\Psi _0^2\left\{
\theta \right\} \left( (a_1a_2)^2+(a_2a_3)^2+(a_3a_1)^2\right)  & \sim  & 
3\left( \frac{2g^2}N\right) ^3\sqrt{\frac \pi \mu }
\end{array}
\end{equation}
Therefore for $<a_1^2+a_2^2+a_3^2>$, which is the long-wavelength
contribution to $g^2a^2<A_\mu ^\alpha A_\alpha ^\mu >$, one obtains%
$$
<a_1^2+a_2^2+a_3^2>=\lim _{\mu \rightarrow 0}\frac 1{-3\mu \ln \mu } 
$$
the same result applying for $\left\langle \left( a_1+a_2+a_3\right)
^2\right\rangle $. Therefore the long wavelength contribution, of the vacuum
background, to these expectation values diverges for all values of the
coupling constant.

These results may now be used to find the effect of the gluon background,
associated to the ground state (\ref{2.4}), on the mass and propagation of
elementary fermion excitations. The eigenvalue equation for the Dirac
Hamiltonian of a fermion minimally coupled to this constant background is%
$$
H_1(p)\psi (p)=\left( -\alpha ^ip_i+g\left( A_0^a+\alpha ^iA_i^a\right) 
\frac{\sigma _a}2+\gamma ^0m\right) \psi (p)=E\psi (p) 
$$
To find the contribution of the background to the fermion mass consider a
Lorenz frame where $\stackrel{\rightarrow }{p}=0$. In this frame, without
loss of generality, one makes $A_0^a=0$ by a gauge transformation and
diagonalizes $A_i^aA_i^a$ by space rotations obtaining, as above%
$$
A_1^\alpha =\frac 2g\left( A_1,0,0\right) \textnormal{ ; }A_2^\alpha =\frac
2g\left( 0,A_2,0\right) \textnormal{ ; }A_3^\alpha =\frac 2g\left( 0,0,A_3\right) 
$$
With these choices the eight eigenvalues of $H_1$ are%
$$
\pm \sqrt{m^2+\left( A_1\pm A_2\pm A_3\right) ^2} 
$$
for all possible sign choices. Hence the background contribution to the
squared mass of the single fermion excitations is $\left\langle \left(
A_1+A_2+A_3\right) ^2\right\rangle $ , the same for all eigenstates because
of the symmetry of the state (\ref{2.4}). As seen above, this vacuum
expectation value diverges for all values of the coupling constant.
Therefore single fermion excitations cannot propagate in this background.

For a composite state with overall neutral color however, the total
contribution of the constant background vanishes and its effect on the mass
can only come from background modifications of the many-body interactions.
For a fermion-antifermion state the total energy is\cite{Ito}%
$$
H_1(p_1)+H_2(p_2)+V 
$$
$V$ stands for the two-body interactions and $H_2$ acts on the adjoint
fermions by%
$$
H_2(p)\stackrel{\_}{\psi }^{\dagger }(p)=\left( -\alpha ^ip_i-g\left(
A_0^a+\alpha ^iA_i^a\right) \frac{\sigma _a}2+\gamma ^0m\right) \stackrel{\_%
}{\psi }^{\dagger }(p) 
$$
Hence, the leading order effect of the constant background cancells for the
color neutral combination $\stackrel{\_}{\psi }(p_2)\psi (p_1)$.

\section{Group structure of the QCD\ vacuum with a SU(3) gluon condensate}

In Sect.1 I have already explained why, in the SU(3) case, with a gluon
condensate, the vacuum structure is not uniquely defined by the SU(3)
symmetry, the possible vacuum structures corresponding to representations of
the homogeneous coset space SO(8)/SU(3). This SO(8) being the group of
rotations in the octet representation of SU(3), the imbedding of SU(3) in
SO(8) is uniquely defined (see the Appendix). SU(3) is the dynamical
invariance of the theory that we start with, therefore it is this group that
should control all the dynamical features of the theory: selection rules,
mass matrix structure, etc. The extra SO(8) symmetry emerges only as a
consequence of the realization of the vacuum through the composite
condensates $\left\langle A_a^\mu A_{a\mu }\right\rangle $ and $\left\langle
F_a^{\mu \nu }F_{a\mu \nu }\right\rangle $. Therefore the only place where
the SO(8) extra quantum numbers naturally appear is on the labelling of the
vacuum classes, which may however be equivalent from the point of view of
QCD\ interactions.

The next logical step is to classify the possible vacuum condensate
structures through the irreducible representations of SO(8) and their
reduction under SU(3). The lowest dimensional representations of SO(8) are
the trivial one and those associated to the fundamental weights $\Lambda _1$%
, $\Lambda _2$, $\Lambda _3$ and $\Lambda _4$ \cite{Cornwell}. The trivial $%
[0,0,0,0]$ representation would corresponds to a vacuum without a condensate
structure and therefore is of no interest. $\Lambda _2=[0,1,0,0]$ is the
28-dimensional adjoint representation which under the SU(3) color subgroup (%
\ref{B.3}) reduces into $10+\overline{10}+8$. The most interesting case
corresponds to the other three representations $\Lambda _1=[1,0,0,0]$, $%
\Lambda _3=[0,0,1,0]$ and $\Lambda _4=[0,0,0,1]$ which are 8-dimensional
representations which are also irreducible octets under this SU(3) subgroup.
They are one-dimensional representations of the homogeneous coset space
SO(8)/SU(3). Therefore one finds 3 identical vacuum structures which are
however distinguished by SO(8) quantum numbers (the maximal weights or the
values of the Casimir operators).

As far as SU(3) color is concerned the three vacuum backgrounds are
equivalent and, if the mass matrix is SO(8)-blind, it is an example of the
democratic mass matrix with all elements equal, discussed by a number of
authors\cite{Fritzsch} \cite{Georgi} \cite{Ramond}. When diagonalized it
leads (in leading order) to one massive state and two massless ones.
Therefore the SO(8) induced background multiplicity is suggestive of a
generation type mechanism. Whether it is really at the origin of the
existence of three particle generations remains to be seen.

\section{Appendix. SO(8) rotations in the octet space}

The antihermitean generators for the Lie algebra of SO(8) have the
commutation relations 
\begin{equation}
\label{B.1}\left[ M_{pq},M_{rs}\right] =\delta _{qr}M_{ps}-\delta
_{qs}M_{pr}-\delta _{pr}M_{qs}+\delta _{ps}M_{qr} 
\end{equation}
$p,q,r,s=1\cdots 8$

The generator $M_{pq}$, with matrix elements 
\begin{equation}
\label{B.2}\left( M_{pq}\right) _{jk}=\delta _{pj}\delta _{qk}-\delta
_{pk}\delta _{qj}
\end{equation}
in the defining representation, generates rotations in the plane $\left(
pq\right) $. The structure constants of a Lie algebra are the matrix
elements of the adjoint representation. Therefore, from the usual structure
constants $f_{abc}$ of SU(3), one reads the representation of the
antihermitean SU(3) generators $F_a$ as functions of the SO(8) rotations in
the octet space%
$$
\left( \frac{\lambda _a}2\longleftrightarrow iF_a\right)  
$$
\begin{equation}
\label{B.3}
\begin{array}{ccl}
F_1 & = & -M_{23}-\frac 12M_{47}+\frac 12M_{56} \\ 
F_2 & = & M_{13}-\frac 12M_{46}-\frac 12M_{57} \\ 
F_3 & = & -M_{12}-\frac 12M_{45}+\frac 12M_{67} \\ 
F_4 & = & \frac 12M_{17}+\frac 12M_{26}+\frac 12M_{35}-
\frac{\sqrt{3}}2M_{58} \\ F_5 & = & -\frac 12M_{16}+\frac 12M_{27}-\frac
12M_{34}+
\frac{\sqrt{3}}2M_{48} \\ F_6 & = & \frac 12M_{15}-\frac 12M_{24}-\frac
12M_{37}-
\frac{\sqrt{3}}2M_{78} \\ F_7 & = & -\frac 12M_{14}-\frac 12M_{25}+\frac
12M_{36}+
\frac{\sqrt{3}}2M_{68} \\ F_8 & = & -\frac{\sqrt{3}}2M_{45}-\frac{\sqrt{3}}%
2M_{67}
\end{array}
\end{equation}
To make the Cartan algebra of SU(3), $\left\{ F_3,F_8\right\} $, a
subalgebra of the Cartan algebra of SO(8) it is convenient to choose this
latter as 
\begin{equation}
\label{B.4}\left\{ h_i\right\} =\left\{ M_{12},M_{45},M_{67},M_{38}\right\} 
\end{equation}
The three octet representations [1,0,0,0], [0,0,1,0] and [0,0,0,1] of SO(8) 
\cite{Cornwell} are also irreducible octets of the SU(3) subgroup defined in
(\ref{B.3}). The correspondence between states is the following: 
\begin{equation}
\label{B.5}
\begin{array}{ccc}
\Lambda _1=[1,0,0,0] &  &  \\  
&  &  \\ 
SU(3) &  & SO(8) \\  
&  &  \\ 
\left| \frac 12\frac 121\right\rangle  & \longleftrightarrow  & \left|
0i00\right\rangle  \\ 
\left| \frac 12\frac{-1}21\right\rangle  & \longleftrightarrow  & \left|
00i0\right\rangle  \\ 
\left| 110\right\rangle  & \longleftrightarrow  & \left| i000\right\rangle 
\\ 
\left| 100\right\rangle  & \longleftrightarrow  & \frac 1{
\sqrt{2}}\left( \left| 000-i\right\rangle -\left| 000i\right\rangle \right) 
\\ \left| 1-10\right\rangle  & \longleftrightarrow  & \left|
-i000\right\rangle  \\ 
\left| 000\right\rangle  & \longleftrightarrow  & \frac 1{
\sqrt{2}}\left( \left| 000-i\right\rangle +\left| 000i\right\rangle \right) 
\\ \left| \frac 12\frac 12-1\right\rangle  & \longleftrightarrow  & \left|
00-i0\right\rangle  \\ 
\left| \frac 12\frac{-1}2-1\right\rangle  & \longleftrightarrow  & \left|
0-i00\right\rangle 
\end{array}
\end{equation}
\begin{equation}
\label{B.6}
\begin{array}{ccc}
\Lambda _3=[0,0,1,0] &  &  \\  
&  &  \\ 
SU(3) &  & SO(8) \\  
&  &  \\ 
\left| \frac 12\frac 121\right\rangle  & \longleftrightarrow  & \left| \frac
i2\frac i2\frac i2
\frac{-i}2\right\rangle  \\ \left| \frac 12\frac{-1}21\right\rangle  & 
\longleftrightarrow  & \left| 
\frac{-i}2\frac i2\frac i2\frac i2\right\rangle  \\ \left| 110\right\rangle 
& \longleftrightarrow  & \left| \frac i2\frac i2
\frac{-i}2\frac i2\right\rangle  \\ \left| 100\right\rangle  & 
\longleftrightarrow  & \frac 1{
\sqrt{2}}\left( \left| \frac i2\frac{-i}2\frac i2\frac i2\right\rangle
+\left| \frac{-i}2\frac i2\frac{-i}2\frac{-i}2\right\rangle \right)  \\ 
\left| 1-10\right\rangle  & \longleftrightarrow  & \left| 
\frac{-i}2\frac{-i}2\frac i2\frac{-i}2\right\rangle  \\ \left|
000\right\rangle  & \longleftrightarrow  & \frac 1{
\sqrt{2}}\left( \left| \frac i2\frac{-i}2\frac i2\frac i2\right\rangle
-\left| \frac{-i}2\frac i2\frac{-i}2\frac{-i}2\right\rangle \right)  \\ 
\left| \frac 12\frac 12-1\right\rangle  & \longleftrightarrow  & \left|
\frac i2
\frac{-i}2\frac{-i}2\frac{-i}2\right\rangle  \\ \left| \frac 12\frac{-1}%
2-1\right\rangle  & \longleftrightarrow  & \left| \frac{-i}2\frac{-i}2\frac{%
-i}2\frac i2\right\rangle 
\end{array}
\end{equation}
\begin{equation}
\label{B.7}
\begin{array}{ccc}
\Lambda _4=[0,0,0,1] &  &  \\  
&  &  \\ 
SU(3) &  & SO(8) \\  
&  &  \\ 
\left| \frac 12\frac 121\right\rangle  & \longleftrightarrow  & \left| \frac
i2\frac i2\frac i2\frac i2\right\rangle  \\ 
\left| \frac 12\frac{-1}21\right\rangle  & \longleftrightarrow  & \left| 
\frac{-i}2\frac i2\frac i2\frac{-i}2\right\rangle  \\ \left|
110\right\rangle  & \longleftrightarrow  & \left| \frac i2\frac i2
\frac{-i}2\frac{-i}2\right\rangle  \\ \left| 100\right\rangle  & 
\longleftrightarrow  & \frac 1{
\sqrt{2}}\left( \left| \frac i2\frac{-i}2\frac i2\frac{-i}2\right\rangle
-\left| \frac{-i}2\frac i2\frac{-i}2\frac i2\right\rangle \right)  \\ \left|
1-10\right\rangle  & \longleftrightarrow  & \left| 
\frac{-i}2\frac{-i}2\frac i2\frac i2\right\rangle  \\ \left|
000\right\rangle  & \longleftrightarrow  & \frac 1{
\sqrt{2}}\left( \left| \frac i2\frac{-i}2\frac i2\frac{-i}2\right\rangle
+\left| \frac{-i}2\frac i2\frac{-i}2\frac i2\right\rangle \right)  \\ \left|
\frac 12\frac 12-1\right\rangle  & \longleftrightarrow  & \left| \frac i2
\frac{-i}2\frac{-i}2\frac i2\right\rangle  \\ \left| \frac 12\frac{-1}%
2-1\right\rangle  & \longleftrightarrow  & \left| \frac{-i}2\frac{-i}2\frac{%
-i}2\frac{-i}2\right\rangle 
\end{array}
\end{equation}
where the SU(3) quantum numbers are $\left| II_3Y\right\rangle $, with $%
I_3=iF_3$ and $Y=i\frac 2{\sqrt{3}}F_8$, and the SO(8) quantum numbers are
the eigenvalues of the antihermitean generators of the Cartan algebra $%
\left\{ h_i \right\} $.

\end{document}